\documentstyle [preprint,aps] {revtex}
\begin {document}
\draft
\title {Intrinsic Pinning in the High Field C-Phase of $UPt_3$}

\author {Brett Ellman and Louis Taillefer}

\address {Department of Physics,
McGill University, 3600 University Street,
Montr\'eal, Qu\'ebec, Canada H3A 2T8}

\date{\today}
\maketitle

\begin {abstract}
We report on the a.c. magnetic response of
superconducting $UPt_3$ in a d.c. magnetic field.
At low fields ($H < H^*$),
the in-phase susceptibility shows a sharp drop at $T_c$
followed by a gradual decrease with decreasing temperature, while
the out-of-phase
component shows a large peak at $T_c$ followed by an unusual broad peak.  As
the B-C phase line is crossed ($H>H^*$), however, both the in-phase and
out-of-phase
susceptibilities resemble the zero-field Meissner curves.  These features are
only observed for relatively large a.c. excitation fields.
We interpret these results in terms of a vortex pinning force which, while
comparatively small in the A/B-phases, becomes large enough to
effectively prevent vortex motion in the C-phase.

\end {abstract}


Thermal and acoustic transport measurements\cite{transport}
of the multi-phase heavy fermion
superconductor $UPt_3$ have proven invaluable in deciphering the structure of
the zero-field superconducting gap.  In particular, the ability of thermal
conductivity and ultrasonic attenuation experiments to couple to the
excitation spectrum while ignoring the superconducting fraction enables one to
measure the momentum space distribution of quasiparticles and thus of the gap
itself.
Similar studies of
the high-field C-phase, however, are greatly complicated by
vortex contributions to the electron and phonon scattering rates,
and the intrinsic
structure and behavior of this phase remain largely unknown.
While not usually viewed as such, an a.c. magnetic susceptibility,
$\chi'(H_{dc},T)+i\chi''(T,H_{dc})$,
measurement for fields $H_{dc}>H_{c1}$ may be
regarded as a transport experiment:
the real component, $\chi'$, probes the ability of vortex
supercurrents to shield the bulk while the imaginary
component, $\chi''$, measures the electrical dissipation due
to the motion of normal state quasiparticles (from vortex cores or regions
of vanishing  gap) in the a.c. magnetic field, much as sound
attenuation
measures the dissipation due to quasiparticle motion due to the moving
lattice.
In this paper we present in-field magnetic susceptibility data on $UPt_3$
in the superconducting state that demonstrate a qualitative difference in the
properties of the high field C-phase and the two low-field phases.
Our data are consistent
with a scenario where the vortex pinning in the C-phase is
sharply higher than that in the low-field regime.  These results, consisting of
both field and temperature sweeps, may also help elucidate behavior
previously noted\cite{tenya} in the literature.

The sample is a single crystal cube of dimensions 1.4 $mm^3$ cut from
a polycrystalline ingot prepared in ultra-high vacuum.  The a.c. susceptibility
was measured at low frequencies (17 to 25 Hz) using
a lock-in amplifier (a Stanford Research SR-850).  As the coils were
uncompensated, a field-dependent/temperature-independent susceptibility has
been subtracted from the data (field and temperature sweeps),
as determined by the condition that $\chi ' (T > T_c) = 0$.
The data are normalized by assuming that $\chi'$ drops by $1/4\pi$ on entering
the superconducting state in zero d.c. field as measured by a run taken
before the superconducting magnet was energized (and thus in the absence of a
residual d.c. field, which may be as large as 400 Oe in the case of the data
of Figs. 1 and 3).
Susceptibility data are given in c.g.s. units throughout.
The maximum possible phase error (which would result in a mixing of
the true in- and -out-of-phase results) is estimated at about 1 degree from
the magnitude of the out-of-phase signal in the normal state. (This estimate
is also consistent with the quoted absolute
phase accuracy of the lock-in amplifier).
Correcting the
data for such a maximal phase
error does not alter any of our conclusions.  Also, data taken at several
different frequencies show that both $\chi '$ and $\chi ''$ scale linearly
with frequency, as opposed to the quadratic scaling of $\chi ''$
expected if it were simply due to,  e.g., capacitive crosstalk of the in-phase
component (with one factor of frequency entering due to the linear growth of
the
in-phase component while another comes in due to the capacitive
coupling itself).
Self-heating effects due to the a.c. field are believed to be small for a
number of reasons.  Firstly, as noted above, no frequency dependence was
observed, eliminating eddy currents in the sample or mount as a heating
source.  Indeed, heating in the normal state is very small since Tc did not
appreciably change as a function of excitation.  Finally, even if heating is
present, it would not lead to the observed a.c. field amplitude dependence
we observe.  We will return to this point below.
As a check on the experiment,  we note that we
have recently measured a different sample from a separate growth
with a different
shape using compensated coils and new electronics (a Linear Research LR-700).
These results are in excellent agreement with those presented here.
A number of experimental
parameters can affect the susceptibility, including
temperature, d.c. field, frequency and
amplitude of the a.c. field, the
directions along which the magnetic fields are applied, and the time period
over which the typically metastable vortex state is measured.   As mentioned,
at the frequencies used in this work, no anomalousus frequency dependence
was seen.
Also, temperature sweeps at half the rate of the data shown here
gave identical results,
indicating that we are not simply measuring the time
(rather than temperature) dependence of the susceptibility\cite{slow_times}.
Furthermore, the field directions were fixed, $H_{dc} || \hat a$ and
$H_{ac} || \hat b$ (and
thus $H_{dc} \perp H_{ac}$).
An important
qualitative dependence on the magnitude of
$H_{dc}$ was observed, however.

In Fig. 1 we show $\chi'(T)$ at various fixed
$H_{dc}$ for $H_{ac} \approx 5 Oe$.  These data, like all the temperature
sweeps shown,
were taken by warming above $T_c$ and then slowly
cooling in field.
The first thing to note is that the susceptibility {\it falls} on
entering the superconducting state.  This is in obvious contrast to an
idealized
(clean)
type II material for which the slope of the magnetization, M(T,H),
is positive\cite{vpin1} at
$H_{c2}$, and is typical of a material with significant vortex
pinning.

Qualitatively, this follows from the fact that the
susceptibility is a direct measure of the derivative,
$4\pi\chi=dM/dH_{ac}=d(B-H_{ac})/dH_{ac}$, of the (time dependent)
magnetization induced by the small a.c. field.
In the presence of pinning, the oscillations of B in time are much smaller
than those of $H_{ac}$:  the vortex lattice cannot "follow" the external field
variations, and the flux is at least partially frozen into the sample.  In the
limit of very strong pinning, $dB/dH_{ac} = 0$ and the
susceptibility is $-1/4\pi$,
as in the Meissner state.  Thus the drop in $\chi'$ near $T_c$ is
qualitatively indicative of the strength of the pinning in the superconducting
state.

The data in Fig. 1 may be naturally divided into two regimes: for
low fields ($H_{dc} <$ 5 kOe), the in-phase susceptibility falls
sharply below $T_c$ to a value significantly less negative than $-1/4\pi$
(indicating relatively weak pinning), has a weak shallow minimum,
and then approaches the Meissner value in
a roughly linear fashion.  For larger d.c. fields, the value
characteristic of very strong pinning is obtained within a small temperature
range below $T_c$.
The question naturally arises as to
whether the distinct change evident for $H_{dc} \approx$ 5 kOe is correlated
with the multiple superconducting phases known to exist in the H-T plane of
$UPt_3$.

We show in Fig. 2a the second critical field as defined by the onset
of the initial, sharp drop in $\chi'$ for the data of Fig. 1 (equivalent
results
are obtained if one uses the imaginary susceptibility).
A definite "kink" is
noted at $H_{dc} \approx$ 5 kOe.
Numerous experiments have shown that this
feature arises from a tetracritical point at which
three superconducting
phases (A,B, and C, as shown schematically in Fig. 2a)
and the normal state meet.
The in-phase susceptibility data of Fig. 1 therefore imply that the vortex
lattice pinning is significantly stronger in the high field C-phase than
in the low-field/low-temperature B-phase\cite{what_about_A}.
Note that this enhanced pinning is
"intrinsic" in the sense that the sample purity, shape, and so on are unchanged
at the B-C
phase boundary.  The strong correlation between the feature in $H_{c2}$ and
the pinning strength is graphically displayed in Fig. 2b, where we
plot the magnitude of the sharp drop in $\chi'$ near $T_c$ as a function of
field:
a precipitous change is noted in the vicinity of the B-C phase boundary.

The data in Fig. 3, measured concurrently with the
$\chi'$ results of Fig. 1, show $\chi''(T)$ for various values of $H_{dc}$.
Three regimes might be posited:  for low fields, a large, sharp, peak is seen
near $T_c$.  However, at intermediate $H_{dc}$, the peak is followed by a
broad second peak extending to the lowest temperatures measured (100mK).
Finally,
for $H_{dc}$ larger than the critical "kink" field of $\approx$5 kOe,
the data once
again resemble the low field results, with a sharp peak near $T_c$ followed
by a monotonic drop.
In interpreting these results, we recall that
the out-of-phase susceptibility, $\chi''$, is a measure of the dissipation
caused by the movement of vortex core
quasiparticles under the influence of
the a.c magnetic field.
If pinning is very strong, the vortices do not
move, and $\chi'' = 0$.
Therefore, the data in Fig. 3 are in qualitative accord with the in-phase
behavior seen in Fig. 1:
vortices are more weakly pinned in the B-phase than in the C-phase.

All of our data show a pronounced peak in $\chi''$ slightly below $T_c$
coincident with the sudden drop in $\chi '$.  We
note that the observation of such a peak is also
seen in some conventional superconductors under the general heading of the
"peak effect," about which a sizable body of literature exists.  In this
work we
mention only that a very sharp peak in $\chi ''$ is also found in a "true"
zero-field run, i.e., one done before the magnet had been ramped, and thus
the
peak is not solely due to the presence of a vortex lattice.  More
interesting from our standpoint is the unexpected presence of a second peak
in $\chi ''$.

The susceptibility was
also found to be a strong function of the strength
of the a.c. magnetic field.  In Fig. 4 we show $\chi(T)$ for two
$H_{dc}$ bracketing the field at the tetracritical point, for various
$H_{ac}$.  At low excitations, the aforementioned features in
$\chi'$ and $\chi ''$ as a function of $H_{dc}$ are absent;  the in-phase
susceptibility drops at $T_c$ to the strong pinning value both above and below
the tetracritical point and the
out-of-phase component exhibits low dissipation in all three phases.
Somewhat similar effects have previously been noted\cite{koziol} in $UPt_3$.
These data emphasize the fact that the phenomena presented here are non-linear
in the sense that they are observed only when vortices are strongly perturbed
about there pinning sites.  One may also see from Fig. 4 that there is a clear
saturation effect:  the curves appear to be reaching an asymptotic shape
with increasing a.c. field amplitude.  We note that this is not what would
be expected from self-heating, which would shift the curves along the
temperature
axis.

Data taken as a function of field allow us to relate the our results
to previous work.  Tenya et
al.\cite{tenya} measured the static magnetization, M,  of $UPt_3$ as a
function of
field for the field along $\hat c$ (orthogonal to our experiment).  They found
a feature in M close to $H_{c2}$, taking the form of a peak or a dip depending
on the field history of the sample.  The feature appears to exist only in the
C-phase.  We have performed a limited number of field sweeps measuring
$\chi(H)$ at fixed T, taken by ramping up the field above $H_{c2}$ and then
stopping at each measurement field on ramping down.

The data at 350mK in Fig. 5 show a sharp drop at
about 5 kOe.  Cuts of temperature sweep data are in relatively good agreement
with the field sweep data.  Comparing the two, we observe that the drop in
5 kOe results from the sharp drop in $\chi'$ at the B-C-phase boundary seen in
Figs. 1 and 2a.  It is tempting to conclude that the feature seen in the
a.c. susceptibility field sweep shown in Fig. 5 is associated with
the d.c. magnetization anomaly
observed near $H_{c2}$ \cite{tenya} by Tenya et al..
However, magnetization computed
in this manner is distinct from that measured in a d.c. experiment, probing as
it does a non-linear response as the field rises and falls.
Furthermore, the d.c. field
direction in the experiment of Tenya et al. is perpendicular to ours.
Thus more work is required to ascertain whether
the features in M(H) observed by Tenya et al. are related to the
sharp changes in pinning observed in this work.
A field dependence of $\chi'$ similar to that of Fig. 5 has also been
seen\cite{upd2al3} in the
heavy fermion superconductor $UPd_2Al_3$.  Whether this is a signature of a
sharp change in vortex pinning (perhaps at a phase line analogous to the B-C
line observed here) remains a subject of future work.

At present, we are not aware of any quantitative theoretical work with
which to interpret our results on the sharp change in pinning in the
C-phase.
To gain some qualitative understanding of what strong
pinning may imply about the Abrikosov lattice, we recall that a vortex is
strongly pinned at a site where the superconducting condensation energy
is small,
since this minimizes the energy gained by the system due to the vanishing order
parameter in the vortex core.  Thus if, for example, the superconducting
order
parameter remains non-zero throughout the vortex lattice, we would expect
relatively weak pinning.  Qualitatively, this is exactly what is predicted
by some particularly successful
theories of the multiple phases of $UPt_3$\cite{vortex_theories}.  The
order parameters of the three phases within these theories are different
linear combinations of two symmetry-allowed functions.
In the B-phase, both components contribute to the order parameter, resulting
in two vortex
lattices spatially separated from each other.  The condensation energy,
calculated as the sum of the condensation energies of the individual
lattices, is everywhere non-zero over the sample.
In the C-phase, however, only one
function is believed to contribute, leading to a conventional vortex
lattice.
This is exactly the scenario envisioned above and results in stronger
pinning in the C-phase than in the B-phase.  Quantitative calculations are
required before the data presented here can be taken as strong evidence for
the exotic inter-penetrating lattice scenario discussed here.  Given the
highly
developed nature of the theories, we hope that the amplitude and temperature
dependence of a.c. measurements may provide quantitative information on
the exotic order parameters in $UPt_3$.

The authors would like to thank Robert Joynt for very useful discussions.
This work was funded by NSERC of Canada and FCAR of Qu\'{e}bec.
L.T. acknowledges
the Canadian Institute for Advanced Research and the A.P. Sloan
Foundation.

\begin{figure}
\caption {In-phase a.c. susceptibility as a function of temperature for various
d.c.
fields with $H_{ac}$=5 Oe.  The zero-field abscissa ranges from 0.51 K to 0.35
K
with the minimum and maximum temperature decreasing by 0.01 K for each
successive field (the exception being the 17 T data which extends from 0.11 K
to 0.27 K).
The drop in $\chi'$ near $T_c$ increases suddenly
around H=5 kOe, indicating strong pinning in the C-phase.  Note also the
similarity of the high field (17 kOe) data with the zero field result.}
\end{figure}

\begin{figure}
\caption{(a) Phase diagram for $UPt_3$ showing multiple
superconducting phases.  The internal phase lines are based on data from
similar samples and are schematic only.  (b) Drop of $\chi'$ near $T_c$ as a
function of field showing the sudden change as the tetracritical point field
is crossed.}
\end{figure}

\begin{figure}
\caption{Out-of-phase a.c. susceptibility as a function of temperature
for various d.c. fields.  The temperature ranges are the same as in Fig. 1.
Two peaks are seen at low d.c. fields, with the broad peak at lower temperature
suppressed for $H_{dc} > $5 kOe.}
\end{figure}

\begin{figure}
\caption{Real and imaginary susceptibility as a function of a.c.
field for d.c. fields above and below the tetracritical point at 5 kOe.
The a.c.
fields range from 0.2 (lower curves) to 4.8 Oe (upper curves) in steps of
1.15 Oe.  The effects indicative of stronger pinning
in the C-phase are only evident at large $H_{ac}$.}
\end{figure}

\begin{figure}
\caption{$\chi'(H)$ at T=350 mK as measured in a field sweep (solid
points) and from isothermal cuts of the data of Fig. 1.  The sharp drop at
$H \approx$ 5 kOe is correlated with the change in $\Delta\chi'$ noted in Fig.
2(b).}
\end{figure}


\begin{references}
\bibitem{vpin1}{This positive susceptibility is very small:
For an extreme type II superconductor like
$UPt_3$, $\chi'_{normal} \approx 1/2\kappa^2 << 1/4\pi$. }
\bibitem{what_about_A}{ The A-phase only exists in a 60 mK interval for H=0 and
vanishes entirely by H $\approx$ 5 kOe.  Thus, strictly speaking, most of
our data only probes the B and C-phases.}
\end{references}
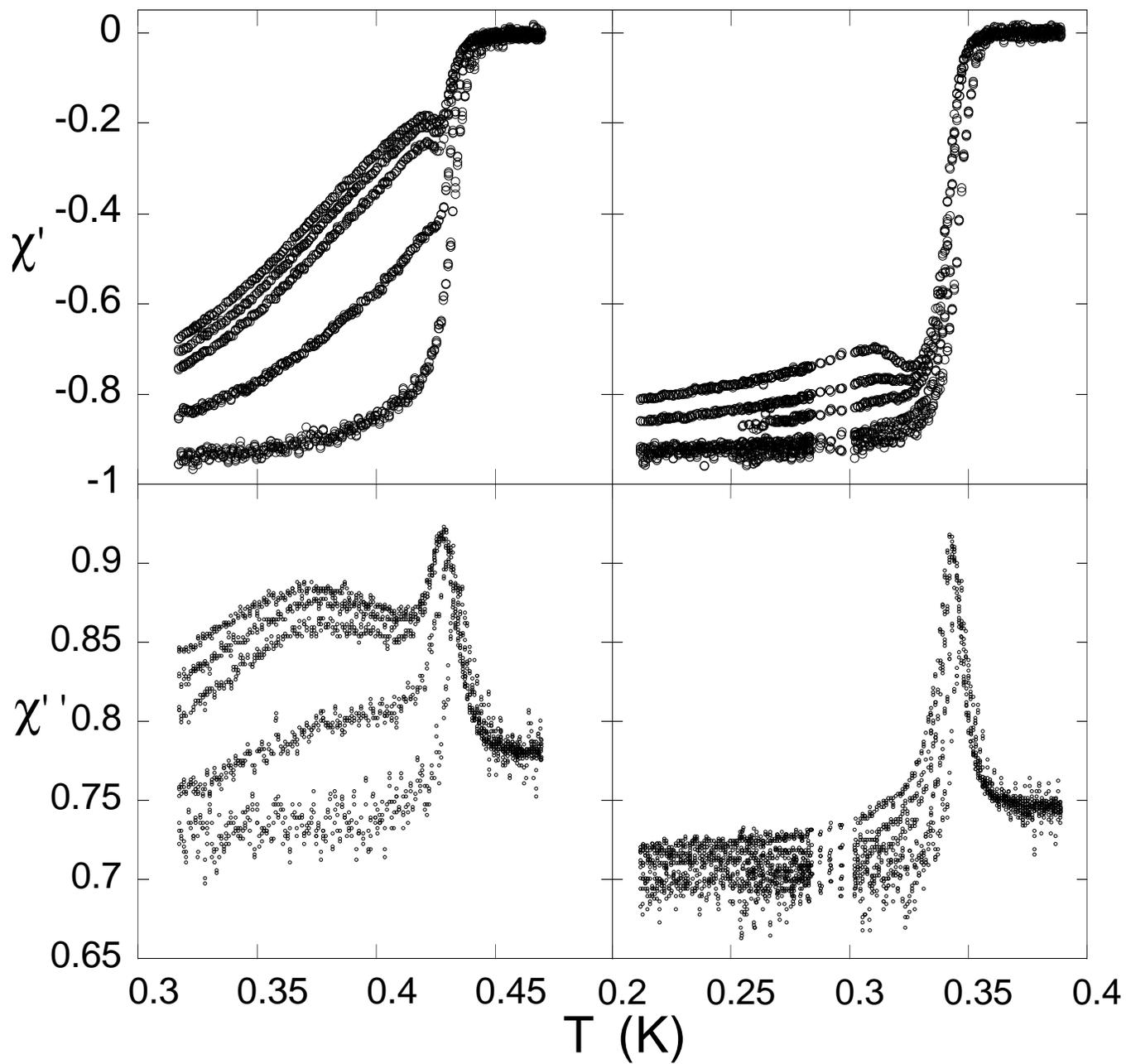
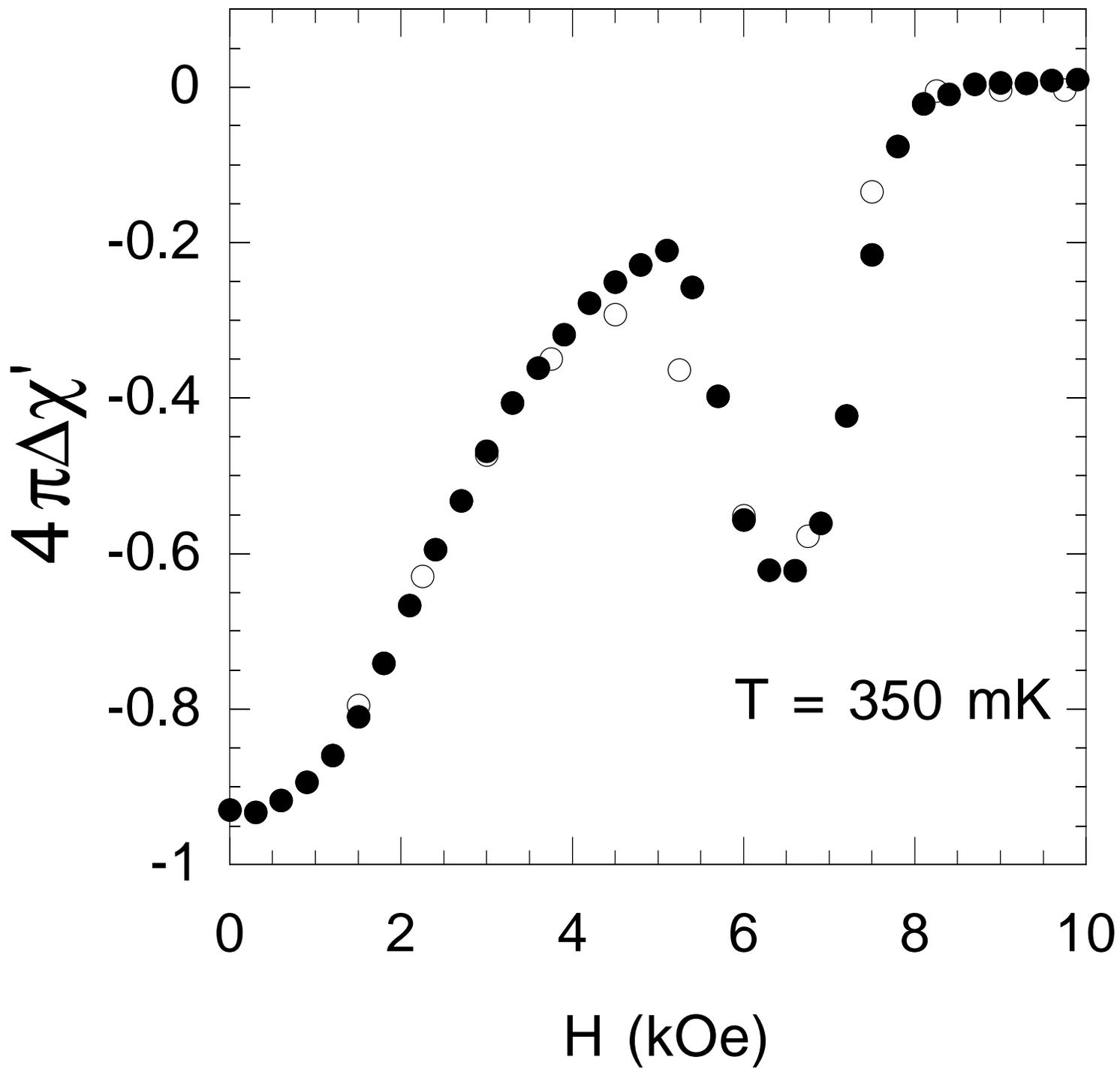
\end{document}